\documentclass[prb,twocolumn,superscriptaddress,showpacs, nofootinbib]{revtex4-2}
\usepackage{graphicx}
\usepackage{float}
\usepackage{dcolumn}
\usepackage{float}
\usepackage{bm}
\usepackage{mathrsfs}
\usepackage{ulem}
\usepackage[per-mode = fraction]{siunitx}
\usepackage{hyperref}
\hypersetup{
    colorlinks=true,
    linkcolor=blue,
    filecolor=magenta,      
    urlcolor=cyan,
    citecolor=magenta,
}
\usepackage{xcolor}
\usepackage{soul}
\usepackage{amsthm,amssymb}
\usepackage{mathtools}
\usepackage{comment}
\usepackage{physics}

\makeatother

\begin{document}
\title{Failure of Mott's formula for the Thermopower in Carbon Nanotubes}
\author{A. V. Kavokin}
\affiliation{Westlake University, Hangzhou 310024, Zhejiang Province, China}
\author{M. E. Portnoi}
\affiliation{Physics and Astronomy, University of Exeter, Stocker Road, Exeter
EX4 4QL, United Kingdom}
\author{A. A. Varlamov}
\affiliation{CNR-SPIN, Via del Fosso del Cavaliere, 100, 00133 Rome, Italy }
\affiliation{Istituto Lombardo ``Accademia di Scienze e Lettere'', via Borgonuovo,
25 - 20121 Milan, Italy}
\author{Yuriy Yerin}
\affiliation{CNR-SPIN, Via del Fosso del Cavaliere, 100, 00133 Rome, Italy }
\date{\today }
\begin{abstract}
Well-known Mott's formula links the thermoelectric power characterised
by Seebeck coefficient to conductivity. 
We calculate analytically the thermoelectric current and Seebeck coefficient in one-dimensional systems and show that, while the prediction of Mott's formula is valid for Dirac fermions, it is misleading for the carriers having a parabolic dispersion. We apply the developed formalism to metallic single wall carbon nanotubes and obtain a non-trivial non-monotonic dependence of the Seebeck coefficient on the chemical potential. We emphasize that, in contrast to Mott's formula, the classical Kelvin's formula that links thermoelectric power to the temperature derivative of the chemical potential is perfectly valid in carbon nanotubes in the ballistic regime. Interestingly, however, the Kelvin's formula fails in two- and three-dimensional systems in the ballistic regime.
\end{abstract}
\maketitle

\section{Introduction}

It is well-known that the low-temperature ballistic conductance in
one-dimensional systems is quantized \cite{sigmaquant}
\begin{equation}
\sigma=\frac{e^{2}J_{\max}}{h},
\label{sigma_J}
\end{equation}
where $h$ is Planck constant while $J_{\max}$ is the number of the
quantization subbands situated below the chemical potential \cite{Wharam, VanWees, vonKlitzing, Tsui82, Ando}. 

Carbon nanotubes (CNTs) represent an example of a one-dimensional system, where indeed
the quantization of conductance has been observed \cite{sigmaCNT}.
It is important to note that the conductance takes constant and discrete
values independently of the type of electronic dispersion: linear
or parabolic. Here we show that in contrast to the conductance
the thermoelectric power in carbon nanotubes is strongly dependent
on the chemical potential, electronic concentration and temperature.
These dependencies are governed by the derivative of the chemical
potential over temperature which is proportional to the Seebeck coefficient $S$
as it was pointed out by Lord Kelvin in the middle of the XIXth
century \cite{Kelvin}
\begin{equation}
S_K=\frac{1}{e}\left(\frac{\partial\mu}{\partial T}\right)_{N,V}
\label{Kelvin}
\end{equation}
(here $N$ is the number of particles, $V$ is the volume of the system).

The alternative, kinetic approach to the theoretical description of the Seebeck effect was developed
by Sir Nevil Mott in the second half of the XXth century \cite{Mott, Heikes}.
The Mott's formula relates the Seebeck coefficient to the conductivity
of the system \cite{Mott_book, Jonson}:
\begin{equation}
S_M=\frac{\pi^{2}}{3}\frac{T}{e}\frac{d\ln\sigma(\mu)}{d\mu},
\label{Mott_1}
\end{equation}
where $\sigma$ is conductivity, $\mu$ is the chemical potential,
$T$ is temperature. Here and further we assume $k_{B}=1.$
One can see that substituting the above expression for conductance
to Mott's formula one obtains $S_M=0$ for any value of chemical
potential except the close vicinity of the bottoms of the quantisation subband. Below we demonstrate that
this simple conclusion fails and Mott's formula is unable to describe
the thermoelectric phenomena in 1D systems in the ballistic regime.
In contrast, the Kelvin's formula remains valid in the ballistic regime in any one-dimensional (1D) system. Interestingly, however, the Kelvin's formula turns out to be no more valid in two- and three-dimensional (2D and 3D) systems, in the ballistic regime. It is worth noting that the Seebeck coefficient can also be introduced through a more general expression for the conductivity within the relaxation time approximation \cite{Ashcroft}. However, we deliberately chose the original definition of Mott formula, which can be more easily extended to the case of the ballistic regime.
We restrict our consideration to the ballistic, semi-classical, linear response regime.
The thermocurrent and thermoelectricity in the ballistic regime have been carefully studied theoretically in 1D, 2D and 3D systems \cite{Sivan, Gurevich, da_Silva}.
The difference between these works and our present work is that we focus on the broken circuit geometry where no electric current is flowing through the system. 
We present an analytical theory of the Seebeck effect in metallic CNTs in the
regime of a ballistic transport. We note that the Seebeck coefficient
is also a direct measure of the entropy per particle, which makes
it one of the most important characteristics of the statistics of
quasiparticles in crystals \cite{Shastry}. We find that, in CNTs, it is dramatically
dependent on the type of electronic dispersion. In the case of a parabolic
dispersion, the Seebeck coefficient is a non-monotonic function of
the chemical potential. Its magnitude decreases with the increase
of the chemical potential, and its sign changes in the vicinity of
the resonances of the chemical potentials and bottom energies of the
electron and hole quantization subbands (the effect of Seebeck coefficient sign change is well established experimental fact often observed in the vicinity of the Fermi surface topology changes \cite{EF82, VY86, Gryaznov82}, see also reviews \cite{Pantsulaya, Yerin} and references therein.). In contrast, in the case of a linear dispersion that is observed in the vicinity of Dirac points in conducting CNTs the Seebeck coefficient is equal to zero. The
sharp contrast between the behaviors of conductivity and Seebeck coefficient
as functions of the chemical potential that may be efficiently controlled
by an applied bias opens room for a variety of non-trivial effects governed by an interplay of currents induced by electric field  and temperature gradient.

\section{Seebeck coefficient in the ballistic regime in a 1D system}
\subsection{1D Kelvin’s formula in the ballistic regime}

The definition of the Seebeck coefficient in the case of a broken electric circuit (${\cal J}=0$, where ${\cal J}$ is an electric current) reads
\begin{equation}
S=\frac{\Delta V}{\Delta T}.
\label{Seebeckdef}
\end{equation}
Here $\Delta T$ is the difference of temperatures at the edges of a $1D$ channel and $\Delta V$ is the voltage induced due to the Seebeck effect. We note that the induction of the voltage ${\Delta V}$ results in the appearance of a ballistic current
\begin{equation}
{\cal J}_V\!=\!e\!\int\displaylimits_{-\infty}^{\infty}\! \nu(E)v(E)\left[f\left(\frac{E\!-\!e \Delta V\!-\!\mu}{T} \right)\!-\!f\left(\frac{E\!-\!\mu}{T}\right)\right]dE,
\label{current_V}
\end{equation}
with $f$ being the Fermi-Dirac distribution, $E$ being the electron energy.
In its turn, the temperature difference $\Delta T$ applied at the edges of the $1D$ channel also will generate the current 
\begin{equation}
{\cal J}_T\!=\!e\!\!\int\displaylimits_{-\infty}^{\infty}\!\nu(E)v(E)\!\left[\!f\left(\frac{E\!-\!\mu(T\!+\!\Delta T)}{T\!+\!\Delta T}\right)\!-\!f\left(\frac{E\!-\!\mu}{T}\right)\right]dE.
\label{current_T}
\end{equation}

The density of states of electrons entering a 1D channel can be written as follows
\begin{equation}
\nu(E)=\frac{1}{2\pi}\left|\frac{\partial E}{\partial k}\right|^{-1}.
\label{DOS_a}
\end{equation}
Here and in what follows we assume the degeneracy factor of quantum states $g=1$. The results can be easily generalized for any other value of $g$.
The electronic velocity is defined by the general relation: 
\begin{equation}
v(E)=\frac{1}{\hbar}\left(\frac{\partial E}{\partial k}\right).
\label{velocity}
\end{equation}
As expected, in every single band their product is constant $\nu(E)v(E)=\frac{1}{h}.$

Note that at negative energies, the Seebeck effect is dominated by holes rather than electrons. Holes are conveniently described by positive effective masses and, consequently, they have positive group velocities. The energy dependence of the hole contribution to the Seebeck effect is symmetric to one of the electron contribution taken with a negative sign as we discuss in Appendix B. We show, in particular, that at the Dirac point, electron and hole contributions to the Seebeck coefficient exactly compensate each other. 

The condition of a broken circuit imposes that the total current in the channel is zero. This condition implies the compensation of the currents (\ref{current_V}) and (\ref{current_T}). 
Expansion of the Fermi-Dirac functions in the limit ${\Delta V}, {\Delta T} \rightarrow 0$ results in the equality
\begin{equation}
\int_{-\infty}^{\infty}\left[-e \Delta V \frac{\partial f}{\partial E} -\Delta T \frac{\partial f}{\partial T} -\left(\frac{\partial\mu}{\partial T}\right)\Delta T \frac{\partial f}{\partial \mu}\right]dE =0.
\label{compensation3}
\end{equation}
The second term is an odd function of energy, hence after integration it yields zero.
In what concerns the partial derivatives of the Fermi-Dirac distribution function over energy and chemical potential, their absolute values are equal while their signs are opposite: ${\partial f}/{\partial E} = -\partial f/{\partial \mu}.$

The remaining integration in Eq. (\ref{compensation3}) is trivial and it leads to 
\begin{equation}
e \Delta V -\left(\frac{\partial\mu}{\partial T}\right) \Delta T =0.
\label{equality3}
\end{equation}
This, in view of the definition (\ref{Seebeckdef}) yields the Kelvin's formula (\ref{Kelvin}).
We conclude that the Kelvin's formula is perfectly valid in any 1D system in the ballistic regime.

Here we would like to point out that in 2D and 3D cases the similar reasoning is also possible, however, the product of the density of states and the electronic velocity appears to be energy dependent. Hence, it cannot be simplified as it was done above. This is why, the equivalent of the second term in Eq. (\ref{compensation3}) would not be equal to zero. An important implication of this is the failure of the Kelvin's formula in 2D and 3D cases in the ballistic regime.

Coming back to the 1D case, at low enough temperatures ($T \ll \mu$) one can rewrite the Kelvin's formula in terms of the logarithmic derivative of DOS (See Appendix \ref{sec:A}). 
\begin{equation}
S_K=\frac{1}{e}\left(\frac{\partial\mu}{\partial T}\right)_{N,V}=\frac{\pi^{2}}{3e}T\frac{\partial}{\partial\mu}\ln\nu(\mu).
\label{muT}
\end{equation}

\subsection{1D Mott’s formula in the ballistic regime}
It is instructive to compare the exact expression for the Seebeck
coefficient obtained above with an approximation that follows from Mott's formula.
In order to derive it, we start from the generic expression for the
electric current density in a one-dimensional channel in the ballistic regime (\ref{current_V}).

We assume that the applied voltage is small, so that $eV\ll T$. For simplicity, in DOS we shall only account for the
contributions of two energy subbands of the CNT. We shall assume that
the chemical potential is close to the bottom of the second subband
from the either positive or negative side. In this case, DOS
may be written as 
\begin{equation}
\nu (E) = \frac{1}{{2\pi }}{\left| {\frac{{\partial E}}{{\partial k}}} \right|^{ - 1}}\left[ {1 + \theta \left( {E - {E_p}} \right)} \right],
\label{DOSdue}
\end{equation}
where $E_{p}$ is the bottom of the second subband. Now one can find
the conductance as
\begin{equation}
\sigma=\frac{\cal J }{V}=\frac{e^{2}}{2h}\left[3-\tanh\frac{E_{p}-\mu}{2T}\right].
\label{sig}
\end{equation}

This expression can be now substituted to
Mott's formula (\ref{Mott_1}). As a result we obtain: 
\begin{equation}
S_M(\mu)=\frac{\pi^{2}}{6e}\frac{1}{\cosh^{2}\frac{E_{p}-\mu}{2T}}\frac{1}{\left(3-\tanh\frac{E_{p}-\mu}{2T}\right)}.
\label{seedue}
\end{equation}
One can see that in the limit of $|E_{p}-\mu|\gg T$ the Seebeck
coefficient calculated in this way appears to be exponentially small. Below we will show
that this prediction of the Mott's formula  strongly deviates from one of the Kelvin's formula confirmed by the exact microscopic consideration (Eqs. (\ref{current_V})-(\ref{equality3})).

\section{The specific case of a metallic CNT}

\subsection{Energy spectra of single-walled carbon nanotubes}

\begin{figure}
\includegraphics[width=0.69\columnwidth]{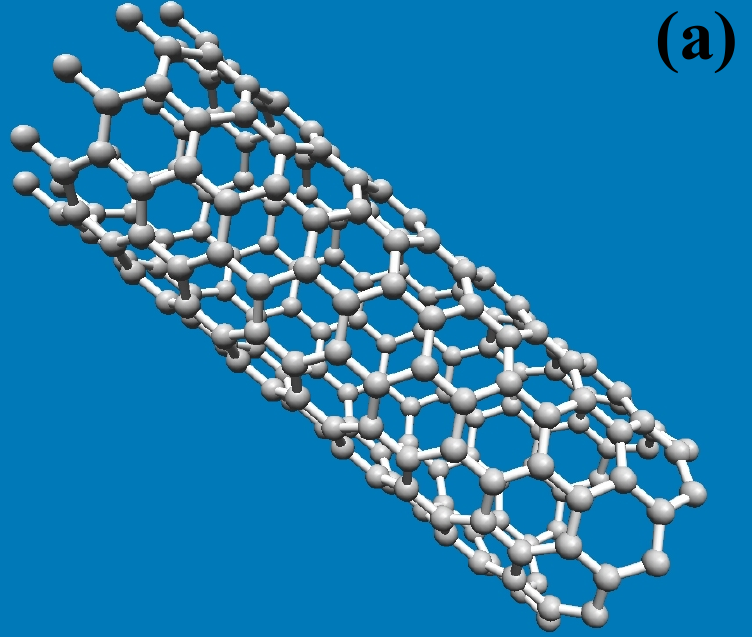}
\includegraphics[width=0.87\columnwidth]{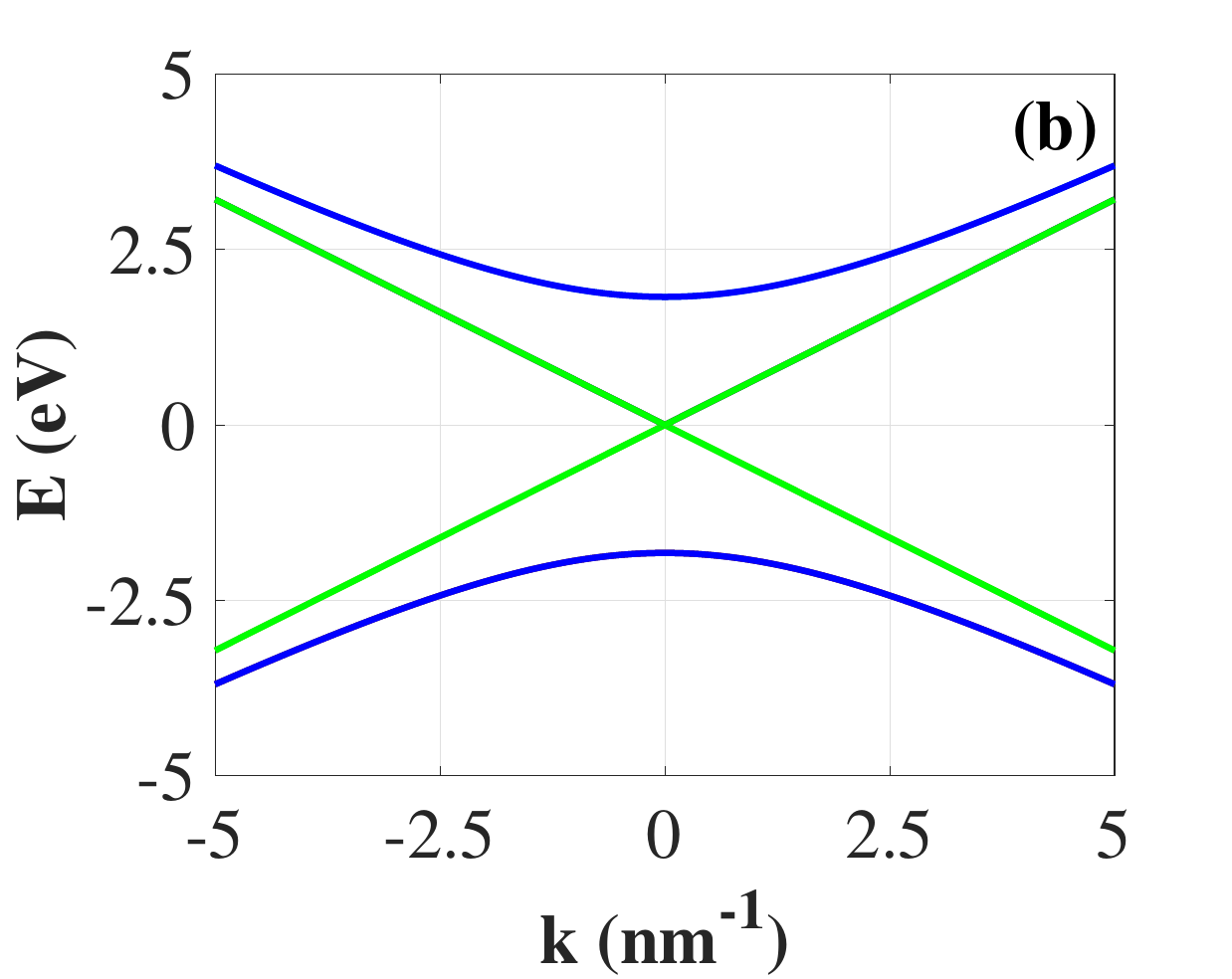}
\caption {(a) A sketch of a zigzag CNT with  chirality $(9,0)$. (b)The energy band structure of a metallic CNT according to the dispersion relation given by Eq. (\ref{CNT_spectrum}) in the limit of zero gap: $E_{0}=0$. Blue lines correspond to the energy levels with a quantum number $j=-1$ and $j=1$, green lines are for the energy levels with $j=0$.}

\label{quasimetallic}
\end{figure}
Basic electronic properties of single-walled (dubbed originally as
single-shell \cite{Iijima93}) carbon nanotubes (CNTs) have been well-understood for over three decades \cite{Saito92} and depend entirely on their chirality $(m,n)$, where (as explained in detail in Ref.~\cite{Saito98}) the numbers $(m,n)$ express the wrapping `chiral' vector connecting
two carbon sites along the nanotube circumference via the two unit
vectors of the underlying hexagonal graphene lattice (see Fig. \ref{quasimetallic}a). A CNT with $n-m\neq3p$,
where $p$ is an integer, is a quasi-one-dimensional semiconductor
with the band gap $E_{g}$ inversely proportional to the nanotube
radius $R=a\sqrt{n^{2}+m^{2}+nm}/(2\pi)\approx0.39\sqrt{n^{2}+m^{2}+nm}$
{\AA}, where the graphene lattice constant $a=\sqrt{3}a_{C-C}$ with $a_{C-C}=1.42$\,{\AA} being a distance between neighbouring carbon atoms in graphene. Only armchair $(n,n)$ nanotubes are truly metallic (gapless) with the low-energy part of dispersion well-described by $E=\pm\hbar v_{F}|k-k_{0}|$, where $v_{F}\approx9.8\times10^{5}$\,m/s is the Fermi velocity in
graphene and $k_{0}=\pm 2\pi/(3a)$.
The remaining CNTs with $n-m=3p$ and $n\neq m$, which are gapless
within the frame of a simple zone-folding approximation \cite{Saito92,Saito98}
of the $\pi$-electron graphene spectrum, are in fact narrow-gap semiconductors
due to the curvature effects~\cite{KaneMele97}.  
For illustration of our main results we use in this work parameters of a (9,0) zigzag nanotube sketched in Fig.~\ref{quasimetallic}(a) assuming it to be metallic with the curvature-induced band gap neglected. The energy spectrum of such a "model" CNT is shown in Fig.~\ref{quasimetallic}(b).

The low-lying branches of CNT spectra, which are relevant for this
work, are well described by cross-sections of the famous graphene
cone for the quantized values of the wave number along the CNT rolling
direction (chiral vector). Thus, in the vicinity of the Dirac point
the energy spectrum of an electron in the $j$-th subband is given
by 
\begin{equation}
E(k)=\pm\sqrt{\hbar^{2}v_{F}^{2}k^{2}+E_{j}^{2}}.
\label{CNT_spectrum}
\end{equation}
For the two branches closest to the Dirac point ($j=0$), for the
metallic CNTs $E_{0}=0$ (see Fig. \ref{quasimetallic}b).
The bottoms of their higher subbands are given by $E_{j}=j\hbar v_{F}/R$. 
Notably, the subbands with $|j|\geq1$  are reasonably-well described
by the effective mass approximation with $m_{j}=E_{j}/v_{F}^{2}$.

\subsection{A vanishing contribution to the Seebeck coefficient from linear dispersion subbands} 

For the linear dispersion subband DOS is constant and given by
$\nu_{0}(\mu)=\frac{2}{\pi \hbar v_{F}}$.
As a result, the temperature derivative of the concentration of charge carriers turns zero. Hence the Dirac mode does not contribute to the Seebeck coefficient in agreement with both the Kelvin's and Mott's formulae. This statement remains valid also in the vicinity of the Dirac point due to the compensation of electron and hole contributions to the Seebeck effect (see Appendix \ref{sec:B}).

\subsection{Conflicting predictions of the Kelvin's and Mott's approaches for the contributions of parabolic subbands to the Seebeck coefficient }
\begin{figure}
\includegraphics[width=1\columnwidth]{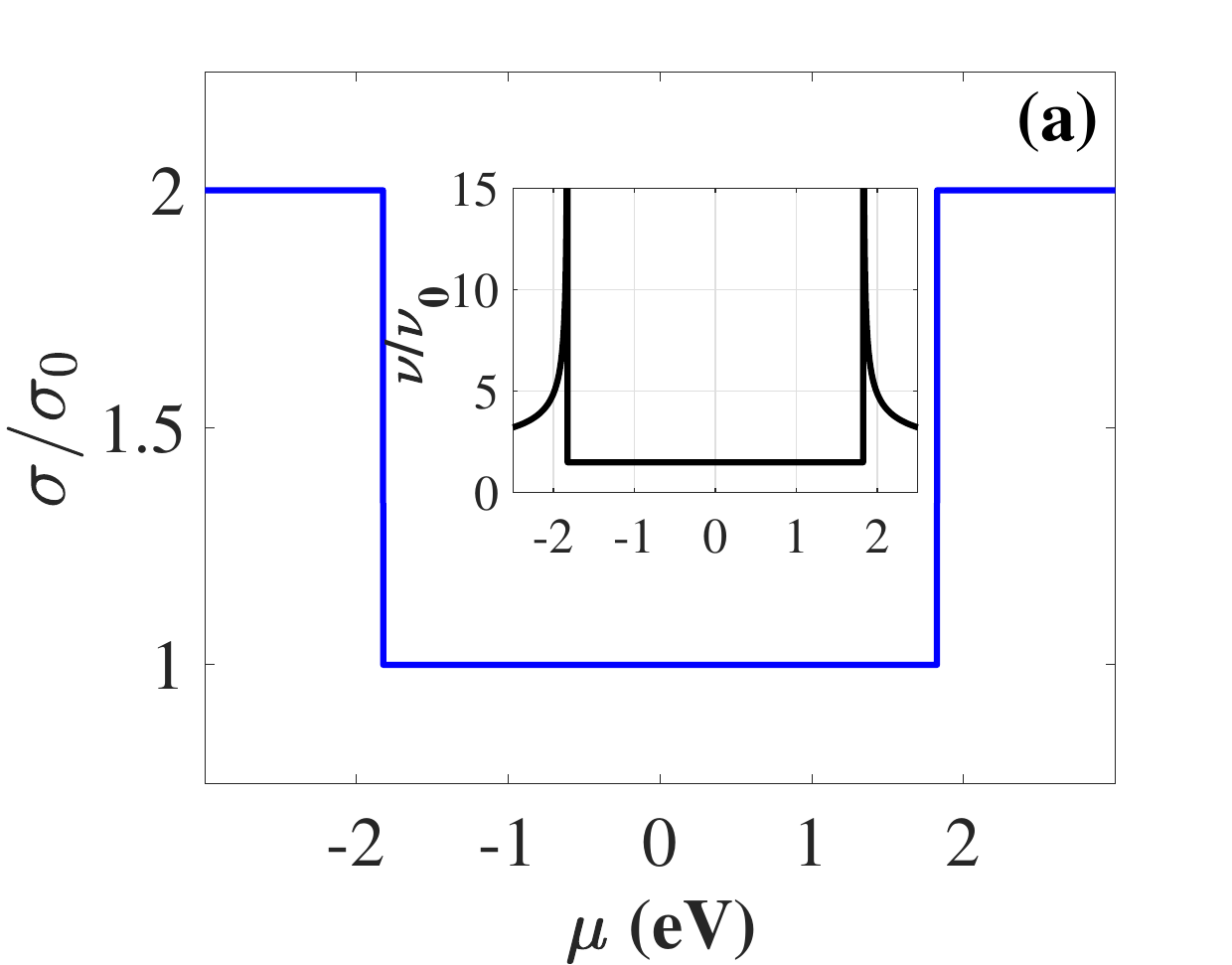}
\includegraphics[width=1\columnwidth]{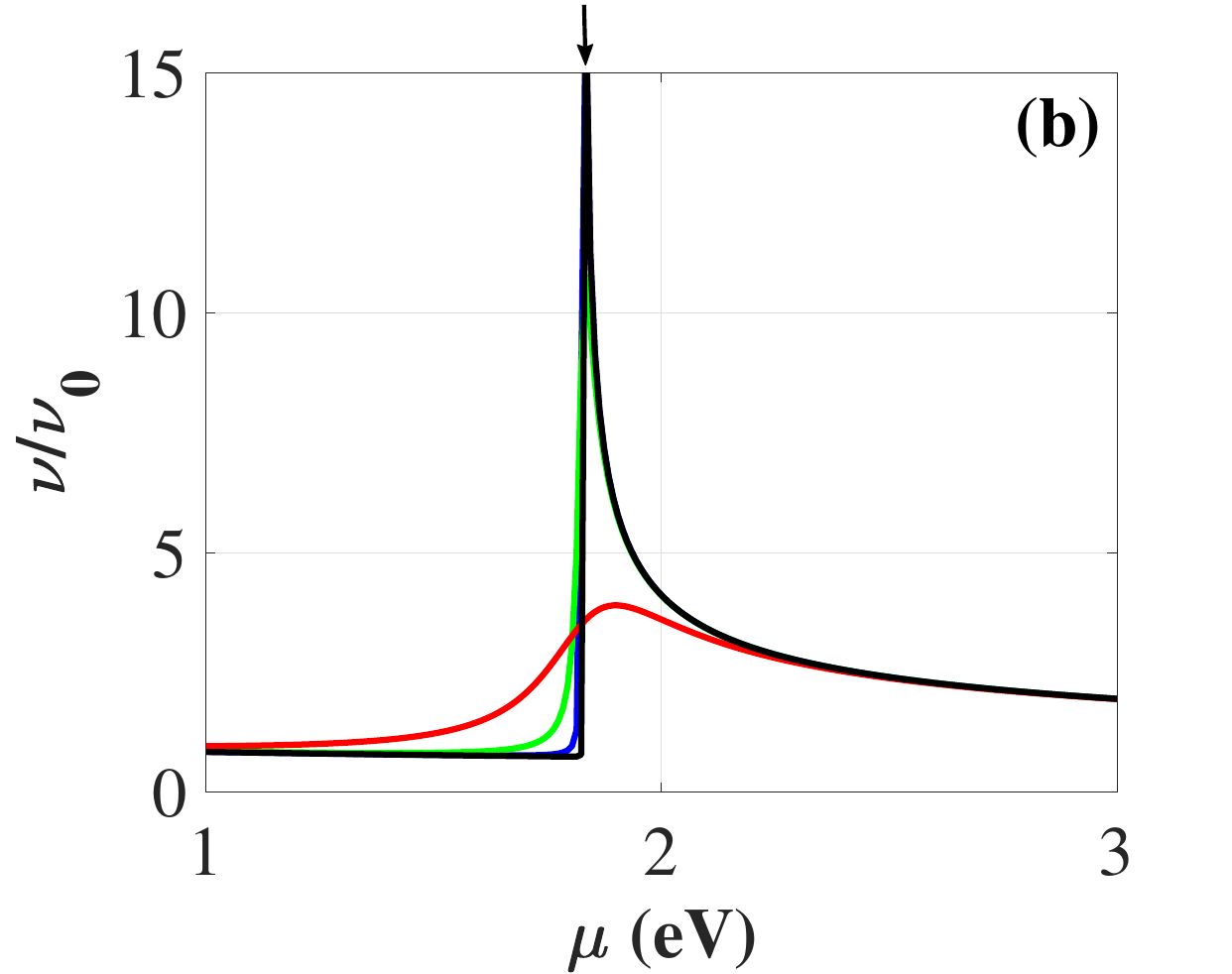}
\caption{(a) The conductance (blue line) for $T=2.4$ K given by Eq.~(\ref{sig}) and the DOS of a metallic CNT without the smearing effect (inset) given by Eq. (\ref{DOS_a}). Here $\nu_0$ is a density of states in the central linear subband and $\sigma_0$ is its contribution to the conductance. (b) DOS of a metallic CNT without the smearing effect (black line) and with smearing one for $\tau/\hbar$=483.09 $eV^{-1}$ (blue line), $\tau/\hbar$=38.65 $eV^{-1}$ (green line) and $\tau/\hbar$=3.86 $eV^{-1}$ (red line). The vertical black arrow at the top of (b) specify the energy corresponding to the bottom of the first subband with $j=1$. Other parameters of a metallic CNT are ${v_F} = 9.8 \cdot {10^5} \rm{m}/\rm{s}$, ${a_c} = 0.142 \, {\rm{ nm}}$ and chirality  $(9,0)$, which gives $R=0.3522 \, {\rm{ nm}}$.}
\label{DOS} 
\end{figure}

Figure \ref{DOS}(a) shows the conductance calculated using Eq.~(\ref{sig}) and the density of states (DOS) calculated with Eq.~(\ref{DOS_a}) for a metallic CNT characterised with the dispersion of electronic bands shown in Fig.~\ref{quasimetallic}(b). The comparison of these curves helps understanding of the qualitative difference between the Seebeck coefficients obtained with the Kelvin’s and the Mott’s formula governed by the derivatives of the DOS and conductivity logarithms over the chemical potential, respectively. Indeed, while DOS in a one-dimensional system is a non-monotonic function of energy, the conductivity is a monotonic staircase-like function. This is why, according to the Kelvin’s formula the Seebeck coefficient changes its sign at each size quantization subband bottom, while according to the Mott’s formula, it remains always positive.

It is also important to note that in all numerical calculations presented in this work we refer to a metallic CNT whose energy spectrum is presented in Fig. \ref{quasimetallic}(b). It contains of the first, linear, band shown by a green line in Fig. \ref{quasimetallic}(b) and the second, parabolic, band shown by the blue line in Fig. \ref{quasimetallic}(b). As we demonstrate analytically, the contribution to the Seebeck coefficient from the linear band is negligibly small at low temperatures. Only the second, parabolic band (and upper bands that are not considered here) gives a sizeable contribution to the Seebeck coefficient. In this context, while one always needs to sum over subbands as it is done in Eq.~(\ref{Seederiv}), for the specific considered case of a metallic CNT, the single band contribution in 
Eq.~(\ref{seedue}) is dominant.

DOS in parabolic subbands can be found as:
\begin{equation}
\nu(\mu)=\sum\limits _{j=1}^{J_{\max}}\frac{\sqrt{2m_{j}}}{\pi \hbar}\frac{1}{\sqrt{\mu-E_{j}}},
\label{DOSparabolic}
\end{equation}
where $J_{\max}$ is the maximum possible value of $j$ such as $\mu-E_{J_{\max}}$ still remains positive.
\begin{figure*}
\includegraphics[width=0.68\columnwidth]{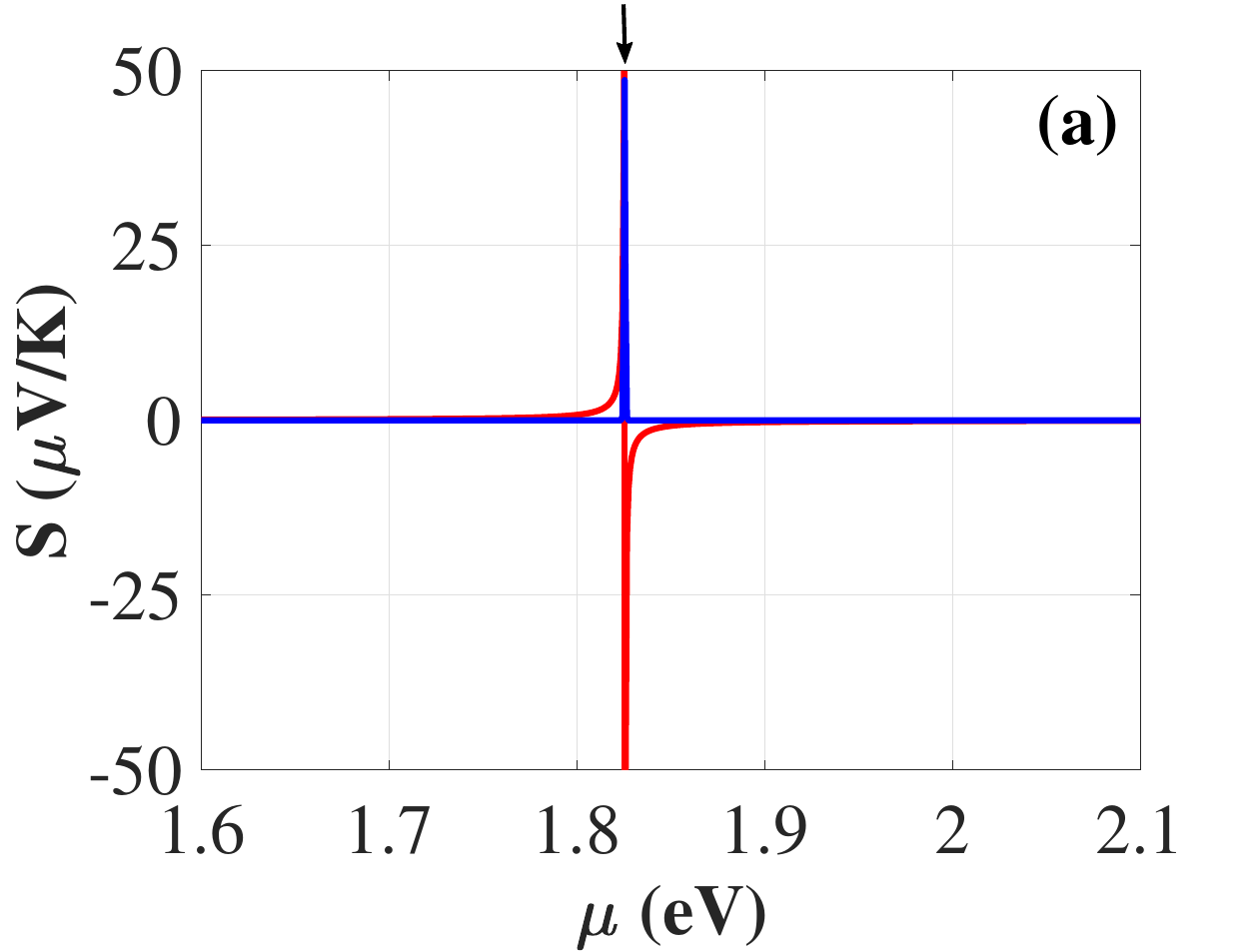}
\includegraphics[width=0.68\columnwidth]{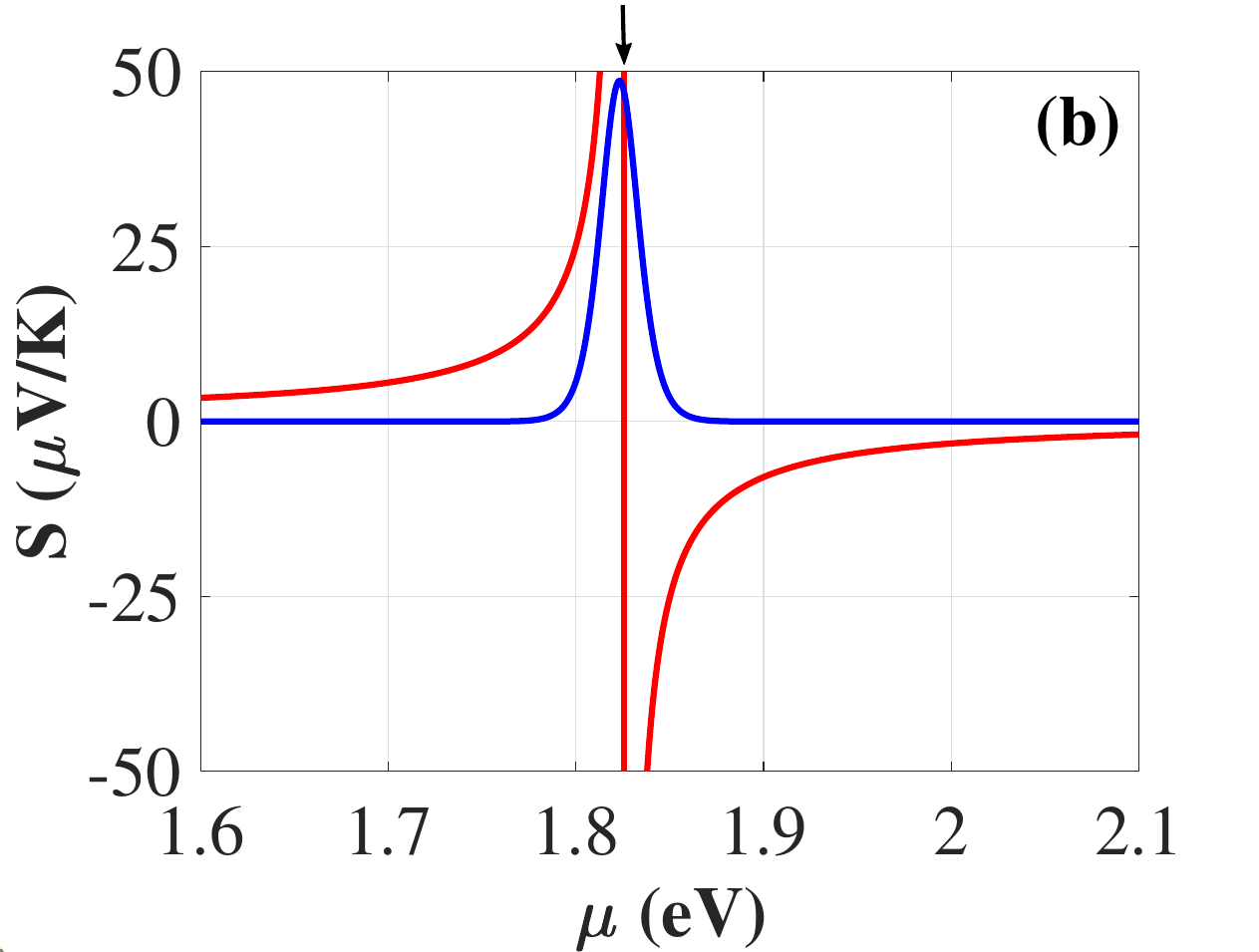}
\includegraphics[width=0.68\columnwidth]{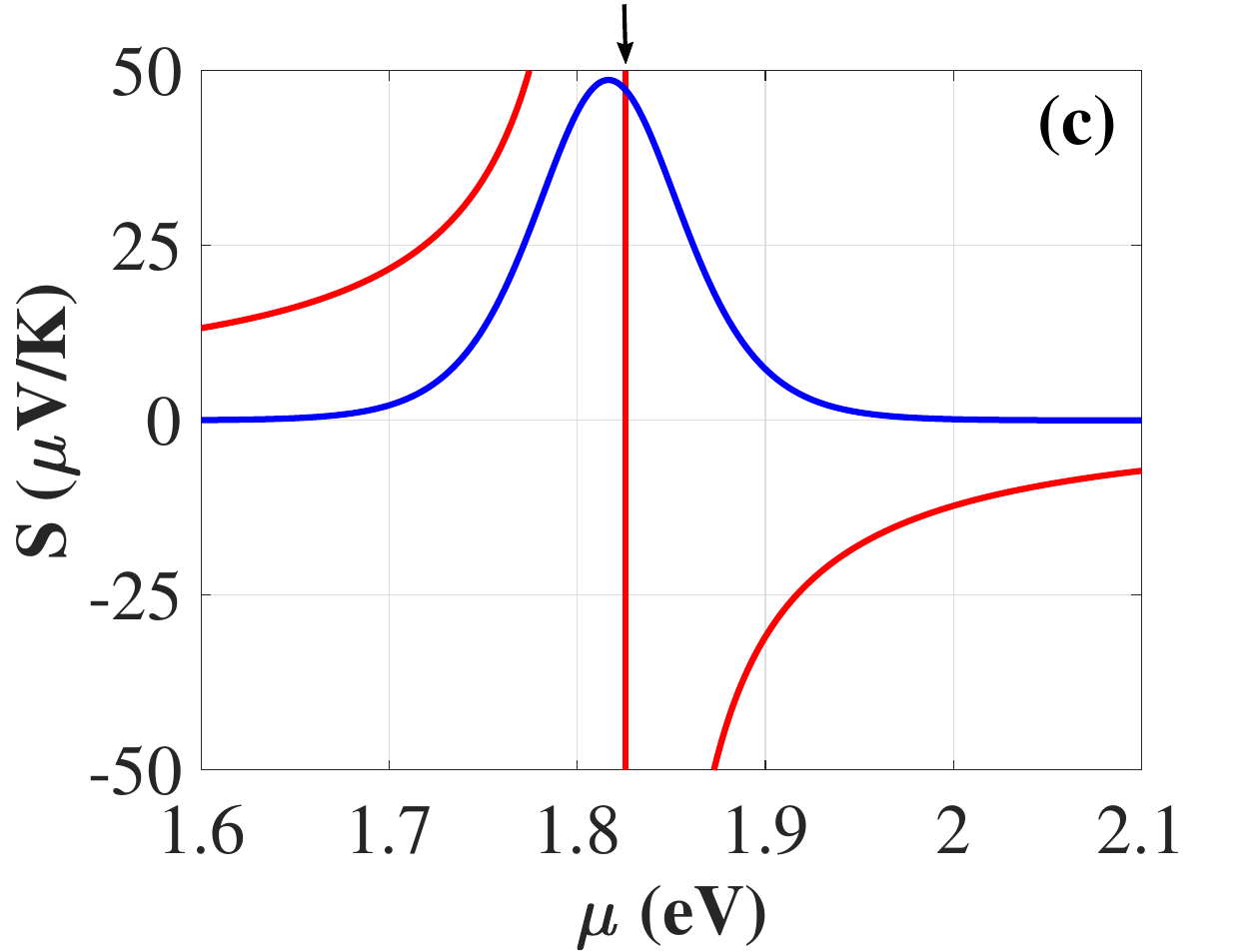}
\caption { The Seebeck coefficient for a metallic CNT as a function of chemical potential $\mu$ calculated for three temperatures T = 2.4 K (a), T = 77 K (b) and T = 300 K (c) according to Kelvin's formula Eq. (\ref{Seederiv}) (red lines) and Mott's formula Eq. (\ref{seedue}) (blue lines). The vertical black arrows at the top of the figures specify the energy corresponding to the bottom of the first subband with $j=1$.}
\label{Seebeck_linear}
\end{figure*}

Using the expressions for DOS and
its derivative one easily finds:
\begin{equation}
S_K(\mu)=\frac{\pi^{2}T}{3eJ_{\max}}\frac{\partial\ln\nu(\mu)}{\partial\mu}=-\frac{\pi^{2}T}{6eJ_{\max}}\frac{\sum\limits _{j=1}^{J_{\max}}\frac{\sqrt{m_{j}}}{\left(\mu-E_{j}\right)^{3/2}}}{\sum\limits _{j=1}^{J_{\max}}\frac{\sqrt{m_{j}}}{\left(\mu-E_{j}\right)^{1/2}}}.
\label{Seederiv}
\end{equation}
The dependence of the Seebeck coefficient on the chemical potential calculated at different temperatures is shown in Fig. \ref{Seebeck_linear}. One can notice the strong difference in the results obtained following the Kelvin's approach (red lines) and Mott's approach (blue lines). The Kelvin's approach predicts the hyperbolic decay of the Seebeck coefficient away from the resonance between the chemical potential and the bottom of the subband. At the resonance point, the singularity and the change of sign of the Seebeck coefficient are observed. In contrast, Mott's formula predicts a finite value of the Seebeck coefficient at the bottom of subband, no sign change and a fast exponential decay away from the resonance. We note at this point that long ago the change of the sign of the Seebeck coefficient was identified as a signature of the topological phase transition which indeed takes place ones the chemical potential crosses the bottom of the next quantization subband \cite{EF82, VP85, VY86, Pantsulaya, Kag94, Pourret17,Pfau17}. The divergence of the Seebeck coefficient in the absence of scattering is caused by the singularity in DOS. It disappears once the smearing of the density of states is taken into account as we show in the next section.

\section{Effect of smearing of DOS in parabolic Subbands}
\subsection{A parabolic subband: the Kelvin's formula}

In any realistic 1D system including CNTs, singularities of DOS are smeared due to a variety of factors: from finite size effects to fluctuations and non-linearities. Therefore, it is important to consider an effect of smearing of DOS on the Seebeck coefficient in the ballistic regime. The Green's function of an electron in a single-wall CNT is
\begin{equation}
G_{j}^{R}(E,p)=\frac{1}{E-p_{z}^{2}/2m_{j}-E_{j}+i \hbar/2\tau},
\label{green}
\end{equation}
where $\tau$ is the positive parameter responsible for the smearing effect and $p_z$ is the continuous momentum along the $z$-direction. This smearing is the result of electron-phonon interaction. 

For the calculation of DOS one can use the standard expression \cite{AGD}:
\begin{equation}
\nu^{(sm)}(E)=-\frac{2}{\pi \hbar}\sum\limits _{j=1}^{J_{\max}}\Im\int\frac{dp}{2\pi}G_j^{R}(E,p).
\end{equation}
Performing the momentum integration and making the correct choice of the branch of
the logarithm one finds
\begin{equation}
\nu^{(sm)}(E)=\sum\limits _{j=1}^{J_{\max}}\frac{\sqrt{m_{j}}}{\pi \hbar }\frac{\sqrt{\sqrt{\left(E-E_{j}\right)^{2}+\frac{\hbar^2}{4\tau^{2}}}+\left(E-E_{j}\right)}}{\sqrt{\left(E-E_{j}\right)^{2}+\frac{\hbar^2}{4\tau^{2}}}},
\label{DOS24}
\end{equation}
where the square root is taken in the arithmetic sense (positive value).
The calculated smeared DOS is presented in Fig. \ref{DOS} .

Now one can derive the contribution to the Seebeck coefficient of a CNT coming from parabolic energy subbands:
\begin{widetext}
\begin{equation}
S^{(sm)}_K(\mu)=\frac{\pi^{2}T}{3eJ_{\max}}\frac{\partial\ln\nu^{(sm)}(\mu)}{\partial\mu}=\frac{\pi^{2}T}{6eJ_{\max}}\frac{\sum\limits _{j=1}^{J_{\max}}\sqrt{m_{j}}\frac{\sqrt{\sqrt{\left(\mu-E_{j}\right)^{2}+\frac{\hbar^2}{4\tau^{2}}}+\left(\mu-E_{j}\right)}}{\left(\mu-E_{j}\right)^{2}+\frac{\hbar^2}{4\tau^{2}}}\left(1-\frac{2(\mu-E_{j})}{\sqrt{\left(\mu-E_{j}\right)^{2}+\frac{\hbar^2}{4\tau^{2}}}}\right)}{\sum\limits _{j=1}^{J_{\max}}\sqrt{m_{j}}\frac{\sqrt{\sqrt{\left(\mu-E_{j}\right)^{2}+\frac{\hbar^2}{4\tau^{2}}}+\left(\mu-E_{j}\right)}}{\sqrt{\left(\mu-E_{j}\right)^{2}+\frac{\hbar^2}{4\tau^{2}}}}}.
\end{equation}
\end{widetext}
The dependence of the Seebeck coefficient on the chemical
potential calculated at different values of smearing is presented in Fig.
\ref{Seebeck_linear}.
\begin{figure}[h]
\includegraphics[width=0.99\columnwidth]{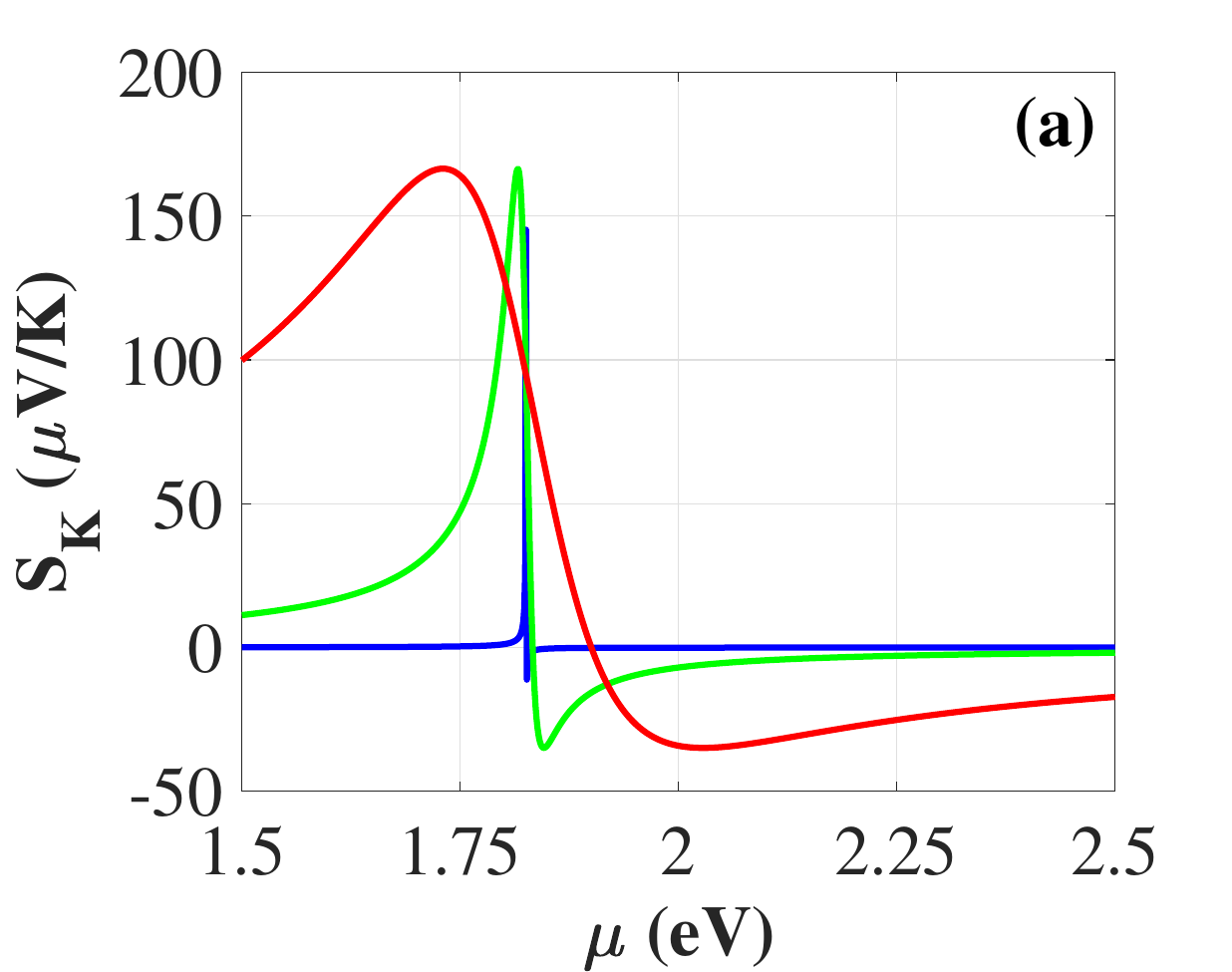}
\includegraphics[width=0.99\columnwidth]{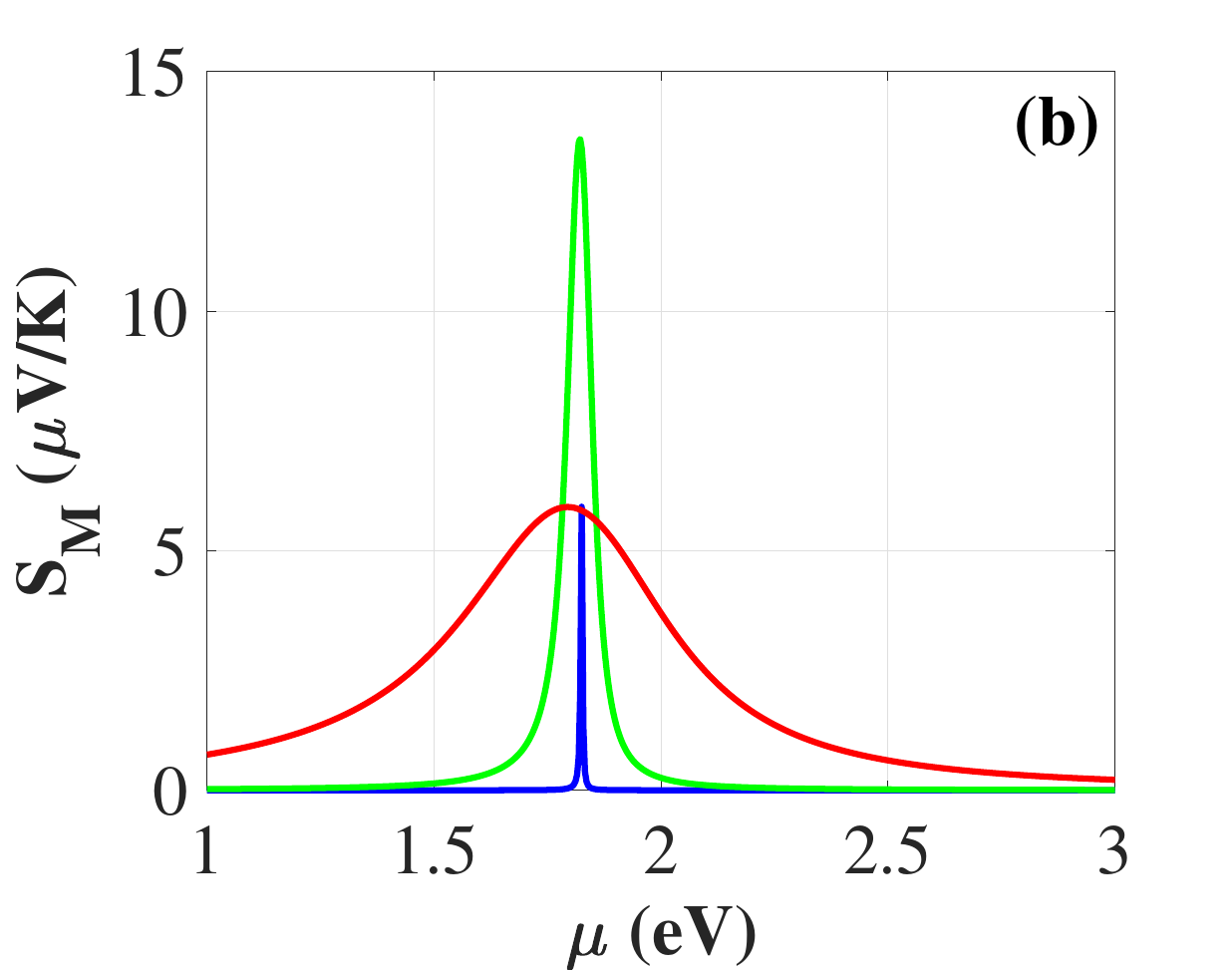}
\caption{The Seebeck coefficient according to Kelvin's formula (a) and Mott's formula (b) as a function of the chemical potential for the parabolic band with the quantum number j=1 for $\tau/\hbar$=483.09 $eV^{-1}$ (blue line), $\tau/\hbar$=38.65 $eV^{-1}$ (green line) and $\tau/\hbar$=3.86 $eV^{-1}$ (red line) .}
\label{Diff_s-1}
\end{figure}

\subsection{A parabolic subband: Mott's formula}
The smearing of the density
of states jump in Eq. (\ref{DOSdue}) will also affect Mott's formula for the Seebeck coefficient. Below we consider a relatively clean system and the range of not too low temperatures: $T\gg\hslash/\tau.$
Within the Lorentz approximation for delta function $\delta\left(E\right)=\frac1{\pi}\lim_{\tau \rightarrow \infty}\hbar\tau^{-1}/\left(E^{2}+\hbar^2\tau^{-2}\right)$ and using the
relation $\delta\left(E\right)=\theta^{\prime}\left(E\right)$, one can
present $\theta$-function in Eq. (\ref{DOSdue}) in the form
\begin{equation}
\theta\left(E\right)=\frac{1}{2}+\frac{1}{\pi}\arctan\left(E \hbar^{-1}\tau\right)\label{theta}.
\end{equation}
This allows to account in Eq. (\ref{seedue}) for the smearing of DOS.
At the same time, temperatures are supposed to be low enough to exclude mixing of electrons belonging to different subbands.

The conductance is no more given by Eq. (\ref{sig}). Instead, it
acquires the form
\begin{equation}
{\sigma ^{(sm)}} \! = \! \frac{{{e^2}}}{{4h}}\frac{1}{{4T}}\int_{ - \infty }^\infty  {\left[ {3 \! + \! \frac{2}{\pi }\arctan \left( {\frac{{\left( {E \! - \! {E_p}} \right)\tau }}{\hbar }} \right)} \right]} \frac{{dE}}{{{{\cosh }^2}\frac{{E \! - \! \mu }}{{2T}}}}.
\nonumber
\end{equation}
Integration of the second term by parts followed by the application of the Cauchy's theorem (see Appendix \ref{sec:C}) yields the conductance explicitly:
\begin{equation}
\sigma^{(sm)}\!=\!\frac{{{e^2}}}{{4h}}\left[ {3\!-\! \frac{2}{\pi }{\mathop{\rm Im}\nolimits} \psi \left( {\frac{1}{2} \! + \! \frac{\hbar}{{2\pi T \tau }} \! + \! i\frac{{{E_p} - \mu }}{{2\pi T}}} \right)} \right].
\label{condimp}
\end{equation}
Substituting it to the Mott's formula one obtains the Seebeck coefficient:
\begin{equation}
S_M^{\left( {{{sm}}} \right)}(\mu) = \frac{1}{{3e}}\,\frac{{{\mathop{\rm Re}\nolimits} {\psi '}\left( {\frac{1}{2} + \frac{\hbar}{{2\pi T\tau }} + i\frac{{{E_p} - \mu }}{{2\pi T}}} \right)}}{{3 - \frac{2}{\pi }{\mathop{\rm Im}\nolimits} \psi \left( {\frac{1}{2} + \frac{\hbar}{{2\pi T\tau }} + i\frac{{{E_p} - \mu }}{{2\pi T}}} \right)}}.
\end{equation}
Here $\psi(x)$ is the digamma-function (logarithmic derivative of the Euler gamma-function) \cite{AS}. 

In the limiting cases where the chemical potential is either close to the bottom of the parabolic subband or sufficiently far from it one finds
\begin{eqnarray}
S_{M}^{(sm)}=	\frac{1}{{3e}}
\begin{cases}
	\frac{1}{3}{\mkern 1mu} {\psi'}\left( {\frac{1}{2} \! + \! \frac{\hbar}{{2\pi T\tau }}} \right) \! + \! \left[\psi'\left( {\frac{1}{2} + \frac{\hbar}{{2\pi  T\tau }}} \right)\right]^2 \cdot \frac{{{E_p} - \mu }}{72\pi ^2T}\\ \qquad \qquad \qquad \qquad \text{when} \qquad |E_{p}-\mu|\ll T,\\
		\frac{{32\pi \hbar }}{11 \cdot T\tau }\frac{{ T^2}}{{{{\left( {{E_p} - \mu } \right)}^2}}} \qquad \text{when} \qquad|E_{p}-\mu|\gg T.
	\end{cases}.
\nonumber
\end{eqnarray}
One can verify that in the limit of $\tau \rightarrow \infty$  this expression reproduces Eq. (\ref{seedue}) \cite{note}. 

Figure \ref{Diff_s-1} (a,b) shows the results of calculation of the Seebeck coefficient as a function of the chemical potential in the vicinity of the bottom of the second electronic subband (j=2) performed accounting for the smearing of DOS. The curves in the panels (a) and (b) are calculated with use of the Kelvin's and Mott's approaches, respectively. One can see that even in the case of a strong smearing Kelvin's and Mott's approaches yield qualitatively different results, especially in what concerns the change of sign of the Seebeck coefficient in the vicinity of the topological phase transition point.

\section{Conclusions}

The ballistic regime offers peculiar modifications of some well-known thermoelectric relations. In particular, it turns out that Mott's formula for the Seebeck effect is incorrect for one-dimensional parabolic bands, while the Kelvin's formula remains fully accurate in this case. This is important also because the application of Kelvin's formula is less challenging numerically than the application of the Mott's formula in many cases. The differences between two approaches become apparent in the vicinity of resonances between the electronic chemical potential and the bottoms of energy subbands characterised by parabolic dispersion. Mott's formula predicts a finite value of the Seebeck coefficient at the resonance and no change of sign. In contrast, the Kelvin's formula predicts divergence and the change of sign of the Seebeck coefficient at the resonance between the chemical potential and the bottom of a parabolic subband. The latter result is characteristic of most of topological phase transitions that occur once a new energy subband comes into play.
Our analysis shows that both Kelvin's and Mott's expressions predict
zero contributions to the Seebeck coefficient from the linear dispersion band in metallic CNTs in the ballistic regime. In contrast, the contribution of parabolic bands is not zero even far from the resonance points, according to the Kelvin's (but not Mott's!) formula. This conclusion can be easily verified experimentally. Any deviation of the Seebeck coefficient of CNTs from zero in the ballistic regime would characterize the inaccuracy of Mott's formula. 
Finally, we note that in 2D and 3D cases, the Kelvin's formula also fails in the ballistic regime. This is because counter-propagating non-dissipative currents of "hot" and "cold" electrons are formed due to the combined actions of the temperature and voltage drops in these cases. The total electric current remains zero, but the system cannot be described by a single electro-chemical potential.
We are looking forward for the experimental manifestations of these theoretical results.

\appendix
\section{The link between the temperature derivative of the chemical potential and the density of states}
\label{sec:A}

We consider the electronic contribution to the Seebeck effect. More general relations accounting for the hole contribution too are given in the Appendix \ref{sec:B}.
To start with, with no loss of generality, the temperature derivative of the chemical potential can be expressed
as follows:
\begin{equation}
\left(\frac{\partial\mu}{\partial T}\right)_{n}=\left(\frac{\partial n}{\partial T}\right)_{\mu}\left(\frac{\partial n}{\partial\mu}\right)_{T}^{-1}.\label{1-2}
\end{equation}
The relationship between the electronic concentration $n$, the chemical
potential $\mu$ and the temperature $T$ can be found by integrating
the density of electron states multiplied by the Fermi-Dirac distribution
over energy: 
\begin{equation}
n\left(\mu,T\right)=\int\limits _{0}^{+\infty}\frac{\nu(E)dE}{\exp\left(\frac{E-\mu}{T}\right)+1}.\label{2-2}
\end{equation}
Required derivatives can be performed assuming $T \ll \mu$
\begin{equation}
\left(\frac{\partial n}{\partial T}\right)_{\mu}=\frac{\pi^{2}T}{3}\frac{d\nu\left(\mu\right)}{d\mu}+\frac{7\pi^4T^3}{90}\frac{d^3 \nu(\mu)}{d\mu^3} +\mathcal{O}
\left(\frac{T^5}{\mu^5}\right),
\label{ngen-1-1-3-1-1}
\end{equation}
and
\begin{equation}
\left(\frac{\partial n}{\partial\mu}\right)_{T}=\nu\left(\mu\right)+\frac{\pi^2T^2}{6}\frac{d^2 \nu(\mu)}{d \mu^2}+ \mathcal{O}
\left(\frac{T^4}{\mu^4}\right).
\label{nmu}
\end{equation}

Hence, in the low temperature limit
\begin{equation}
S_K=\frac 1e \left(\frac{\partial\mu}{\partial T}\right)_{N,V}=\frac{\pi^{2}}{3e}T\frac{d \ln\nu(\mu)}{d\mu}.
\label{muT}
\end{equation}

\section{The vicinity of the Dirac point}
\label{sec:B}

In the vicinity of the Dirac point, both electron and hole concentrations
are different from zero at non-zero temperature. Below we will take
into account the dependence of chemical potential on temperature in
order to evaluate both electron and hole contributions to the Seebeck
coefficient.
\begin{equation}
n_{e,h}\left(\mu,T\right)=\frac{2}{\pi \hbar v_{F}}\int\limits _{0}^{+\infty}\frac{dE}{\exp\left(\frac{E-\mu_{e,h}}{T}\right)+1},\label{neh}
\end{equation}
where we introduced the chemical potential for holes $\mu_{h}=-\mu_{e}=-\mu$. 
Due to the possible nonzero charge of the system, 
\begin{equation}
n_{e}-n_{h}=n_{0}.
\label{n0}
\end{equation}
Here $-|e|n_{0}$ is the overall charge density. Substituting Eq. (\ref{neh}) in  Eq. (\ref{n0}) and performing integration one finds
\begin{equation}
\mu=\mu_{e}=\frac{\pi \hbar v_{F}n_{0}}{2}.\label{mu-n0}
\end{equation}
Thus, one can see that the chemical potential for the linear dispersion
branch depends on the total charge density only and it does not depend
on temperature. This means that $\left({\partial n_{0}}/{\partial T}\right)_{\mu}=0$
and corresponding Seebeck coefficient is zero for the whole linear
branch including the vicinity of the Dirac point: 
\begin{equation}
S_K=\frac{1}{e}\left(\frac{\partial\mu}{\partial T}\right)_{n}=\frac{1}{e}\left(\frac{\partial n_{0}}{\partial T}\right)_{\mu}\left(\frac{\partial n_{0}}{\partial\mu}\right)_{T}^{-1}=0.
\label{S Dirac}
\end{equation}

\subsection{The special point $\mu=0$}

Here we address the special case of the chemical potential resonant with the Dirac point.  We already saw that $\mu_{e}=-\mu_{h}=\mu$ and $n_{e}-n_{h}=n_{0}$. Let us substitute these relations to the above derivatives to the general thermodynamic expression relating the entropy per particle $\left(\partial {\cal S}/\partial n_{0}\right)_{T}$ with partial temperature derivative of the chemical potential
\begin{eqnarray}
&\left(\frac{\partial {\cal S}}{\partial n_{0}}\right)_{T}=\left(\frac{\partial\mu}{\partial T}\right)_{n_{0}}=\left(\frac{\partial n_{0}}{\partial T}\right)_{\mu}\left(\frac{\partial n_{0}}{\partial\mu}\right)_{T}^{-1}= \nonumber \\
&\left[\left(\frac{\partial n_{e}}{\partial T}\right)_{\mu_{e}}-\left(\frac{\partial n_{h}}{\partial T}\right)_{\mu_{h}}\right]\left[\left(\frac{\partial n_{e}}{\partial\mu_{e}}\right)_{T}+\left(\frac{\partial n_{h}}{\partial\mu_{h}}\right)_{T}\right]^{-1}.\label{long}
\end{eqnarray}

One can see that at the point $\mu=0$ the number of electrons coincides with the number of holes ($n_{e}=n_{h}$),  {\it i.e.} the first multiplier in Eq. (\ref{long}) turns zero, while with use of  Eq. (\ref{neh}) it is easy to find that the second multiplier is equal to $\pi v_F/2$. Hence, the entropy per particle and Seebeck coefficient in the Dirac point turn out to be equal to zero:
\[
\left(\frac{\partial {\cal S}}{\partial n_{0}}\right)_{T,\mu=0} = \, eS_K=0.
\]

\section{Derivation of the expression for conductance}
\label{sec:C}

Here we provide the detailed calculation of the integral for conductance:
\begin{equation}
\label{integral_cond}
\sigma  = \frac{{{e^2}}}{{8hT}}\int\limits_{ - \infty }^{ + \infty } {\left[ {3 + \frac{2}{\pi }\arctan \left( {\left( {E - {E_p}} \right) \hbar^{-1} \tau } \right)} \right]\frac{{dE}}{{{{\cosh }^2}\frac{{E - \mu }}{{2T}}}}}.
\end{equation}
We split integral Eq. (\ref{integral_cond}) into parts $\sigma  = {\sigma _1} + {\sigma _2}$, where
\begin{equation}
\label{sigma1}
{\sigma _1} = \frac{{3{e^2}}}{{8hT}}\int\limits_{ - \infty }^{ + \infty } {\frac{{dE}}{{{{\cosh }^2}\frac{{E - \mu }}{{2T}}}}}  = \frac{{3{e^2}}}{{2h}},
\end{equation}
and
\begin{equation}
\label{sigma2}
{\sigma _2} = \frac{{{e^2}}}{{4\pi hT}}\int\limits_{ - \infty }^{ + \infty } {\frac{{\arctan \left( {\left( {E - {E_p}} \right) \hbar^{-1} \tau } \right)dE}}{{{{\cosh }^2}\frac{{E - \mu }}{{2T}}}}}.
\end{equation}

Applying integration by parts to Eq. (\ref{sigma2}) one can rewrite the integral as
\begin{equation}
{\sigma _2} =
- \frac{{{e^2}T\tau }}{{\pi h}}\int\limits_{ - \infty }^{ + \infty } {\frac{{\tanh z dz}}{{1 + {{\left( {2T \hbar^{-1} \tau } \right)}^2}{{\left( {z - \frac{\left(E_{p}-\mu\right)}{2T}} \right)}^2}}}},
\nonumber
\end{equation}
where we introduced new variable $z = \frac{{\left( {E - {E_p}} \right) \hbar^{-1} \tau + \delta }}{{2T\tau }}$ with the parameter $\delta  = \left( {{E_p} - \mu } \right) \hbar^{-1} \tau $. The latter integral can be evaluated using the Cauchy's theorem by means of residues. The appearing summation over the poles of $\tanh z$ can be performed in terms of digamma-function and after straightforward calculations one finds:
\begin{equation}
\label{sigma2_final}
{\sigma _2} = 
  - \frac{{{e^2}}}{{\pi h}}{\mathop{\rm Im}\nolimits} \psi \left( {\frac{1}{2} + \frac{\hbar}{{2\pi T\tau }} + i\frac{{{E_p} - \mu }}{{2\pi T}}}\right),
\end{equation}
where the symmetry property of the digamma function $\psi \left( {\bar z} \right) = \overline {\psi \left( z \right)} $ was applied \cite{AS}. The similar structure of the expression for conductance in terms of the polygamma function was also obtained in Refs. \onlinecite{Pickem1, Pickem2, Valli, Pickem3}.

Therefore, the final result for conducance is expressed as 
\begin{equation}
\sigma \! =\sigma^{(sm)}\! = \! \frac{{{e^2}}}{{2h}}\left[ {3 \! - \!\frac{2}{\pi }{\mathop{\rm Im}\nolimits} \psi \left( {\frac{1}{2} \! + \! \frac{\hbar}{{2\pi T\tau }} \! + \! i\frac{{{E_p} - \mu }}{{2\pi T}}} \right)} \right].
\nonumber
\end{equation}

In the case of $\tau  \to \infty $  using the relation between the imaginary part of the digamma function and the hyperbolic tangent \cite{AS} one can reduce 
Eq.~(\ref{sigma2_final}) to the expression
\begin{equation}
{\sigma _2} \! = \! - \frac{{{e^2}}}{{\pi h}}{\mathop{\rm Im}\nolimits} \psi \left( {\frac{1}{2} \! + \! i\frac{{{E_p} - \mu }}{{2\pi T}}} \right)
 = \! - \frac{{{e^2}}}{{2h}}\tanh \left( {\frac{{{E_p}  \!- \! \mu }}{{2T}}} \right).
\nonumber
\end{equation}
and reproduce  Eq. (\ref{sig}).


\begin{thebibliography}{100}
\bibitem{note} We note here that dealing with the Taylor expansion of the polygamma-functions we were not able to reproduce exponentially small asymptotics of  Eq. (\ref{seedue}) far from the bottom of the parabolic band (when $|E_{p}-\mu|\gg T$), but simply obtained zero.
\end{thebibliography}
\end{document}